\documentclass[aps,pra,twocolumn,showpacs,preprintnumbers,amsmath,amssymb]{revtex4-2}
\usepackage{soul}
\usepackage{amsmath}
\usepackage{esvect}

\usepackage{natbib}
\usepackage{graphicx}
\usepackage{color}
\usepackage{tabularx}

\usepackage{amssymb}

\usepackage[utf8]{inputenc}

\usepackage{float}

\usepackage{subfigure}

\usepackage{float}

\usepackage{hyperref}

\usepackage{cancel}

\usepackage{setspace}

\begin{document}

\title{Intensity nonlinearity of the error-signal frequency shift in the modulation spectroscopy of dark resonances and approaches to its reduction}

\author{E.\,A.~Tsygankov$^{1}$}
\email[]{selentinthebright@gmail.com}
\affiliation{1. The Lebedev Physical Institute of the Russian Academy of Sciences, Moscow, 119991 Russia\\
2. HSE University, Moscow 101000, Russia\\
3. National Research Nuclear University MEPhI, Kashirskoye Highway 31, Moscow, 115409 Russia}

\author{S.\,V.~Petropavlovsky$^{2}$}
\affiliation{1. The Lebedev Physical Institute of the Russian Academy of Sciences, Moscow, 119991 Russia\\
2. HSE University, Moscow 101000, Russia\\
3. National Research Nuclear University MEPhI, Kashirskoye Highway 31, Moscow, 115409 Russia}

\author{M.\,I.~Vaskovskaya$^{1}$}
\affiliation{1. The Lebedev Physical Institute of the Russian Academy of Sciences, Moscow, 119991 Russia\\
2. HSE University, Moscow 101000, Russia\\
3. National Research Nuclear University MEPhI, Kashirskoye Highway 31, Moscow, 115409 Russia}

\author{D.\,S.~Chuchelov$^{1}$}
\affiliation{1. The Lebedev Physical Institute of the Russian Academy of Sciences, Moscow, 119991 Russia\\
2. HSE University, Moscow 101000, Russia\\
3. National Research Nuclear University MEPhI, Kashirskoye Highway 31, Moscow, 115409 Russia}

\author{S.\,A.~Zibrov$^{1}$}
\affiliation{1. The Lebedev Physical Institute of the Russian Academy of Sciences, Moscow, 119991 Russia\\
2. HSE University, Moscow 101000, Russia\\
3. National Research Nuclear University MEPhI, Kashirskoye Highway 31, Moscow, 115409 Russia}

\author{V.\,V.~Vassiliev$^{1}$}
\affiliation{1. The Lebedev Physical Institute of the Russian Academy of Sciences, Moscow, 119991 Russia\\
2. HSE University, Moscow 101000, Russia\\
3. National Research Nuclear University MEPhI, Kashirskoye Highway 31, Moscow, 115409 Russia}

\author{V.\,L.~Velichansky$^{1,3}$}
\affiliation{1. The Lebedev Physical Institute of the Russian Academy of Sciences, Moscow, 119991 Russia\\
2. HSE University, Moscow 101000, Russia\\
3. National Research Nuclear University MEPhI, Kashirskoye Highway 31, Moscow, 115409 Russia}

\author{V.\,P.~Yakovlev$^{3}$}
\affiliation{1. The Lebedev Physical Institute of the Russian Academy of Sciences, Moscow, 119991 Russia\\
2. HSE University, Moscow 101000, Russia\\
3. National Research Nuclear University MEPhI, Kashirskoye Highway 31, Moscow, 115409 Russia}

\begin{abstract}
\vspace{0.5cm}
We have found that the error-signal frequency corresponding to the coherent population trapping resonance can be displaced from that of ``$0-0$'' transition unperturbed by the optical field, although the frequency is not sensitive to changes in its intensity. We consider the double $\Lambda$-system of levels interacting with the asymmetric polychromatic optical field to demonstrate that this effect arises due to intensity nonlinearity of the error-signal frequency shift. The experiment with $^{87}$Rb atoms in Ar-N$_2$ buffer gas atomic cell shows how the displacement value depends on different parameters. The possible influence of the effect on the clocks' frequency stability and reproducibility are discussed.
\vspace{1.5cm}
\end{abstract}

\maketitle

\section{Introduction}

Compact coherent population trapping (CPT) clocks~\cite{doi:10.1063/1.1691490,doi:10.1063/1.1787942,doi:10.1063/1.5026238} are rapidly developed devices since they provide a possibility to achieve higher long-term frequency stability compared with quartz oscillators alongside with drastic reduction of the microwave standards size~\cite{doi:10.1063/1.1494115}. The negative concomitant effect of the all-optical pumping and interrogation scheme utilizing the CPT phenomenon is the light shift~\cite{refId0,PhysRev.171.11,PhysRevA.5.968,PhysRevA.27.1914,PhysRevA.58.2345,0295-5075-48-4-385,Wynands1999}, leading to a frequency drift of the local oscillator (LO) (if the optical field intensity changes), which limits the long-term frequency stability~\cite{Knappe:01,Vanier2005,Gerginov:06,SHAH201021,2015NCimR..38..133G}. One of the standard approaches of the CPT resonances excitation is the usage of a polychromatic field obtained by modulation of the injection current of the vertical-cavity surface-emitting laser~(VCSEL)~\cite{michalzik2012vcsels,Affolderbach2000}. In many papers it was demonstrated that in the case of such fields one can get rid of the light shift by a proper choice of the depth of the laser injection current modulation or by a proper choice of the phase modulation index, if, for example, the electro-optic modulator is used~\cite{827437,articleZhu,articleLutwak,Affolderbach2005,doi:10.1063/1.2360921,Mikhailov:10,doi:10.1063/1.3530951,Miletic2012,Zhang:16}. This is simply because the higher-order spectrum components induce the non-resonant light shift of the opposite sign than the carrier and resonant (first-order) sidebands. Consequently the compensation of partial light shifts is possible (it refers to the case where the modulation frequency is equal to the half of the frequency of an alkali metal atom ground-state hyperfine interval).

It was supposed in quite a number of the cited above works, that for different spectra of an optical field providing the zero light shift, the CPT resonance frequency is the same. Namely, the two-photon detuning in such a case is equal to zero, i.e., the resonance frequency is equal to that of the ``0-0'' transition unperturbed by optical fields. However, in our recent study of the buffer gas effect on the light shift suppression possibility, we obtain the other result~\cite{BGpaper}. The experiment was performed with the modulated VCSEL and $^{87}$Rb atoms, and the CPT resonance frequency was tracked employing the technique of the modulation spectroscopy. We have observed vanishing of the frequency sensitivity to variations of the total laser intensity for two modulation depths (i.e., for two different spectra). Frequencies of the error-signal zero-crossing point were different in the case of insensitivity.

In this work we demonstrate that this difference can occur at high values of the modulation frequency (which provides the error signal) in the case of unequal powers of the resonant spectrum components
and also due to the optically thick medium. In these cases, the error-signal frequency shift becomes a nonlinear function of the laser power, which gives two different conditions for spectra providing the insensitivity and suppression of the light shift. We will refer to these cases as the insensitivity point (IP) and the point of the zero displacement (PZD), respectively. 
To demonstrate the divergence analytically, we consider the optical field with unequal resonant sidebands and phenomenologically take into account their absorption over the length of an atomic cell. The density matrix approach is used to obtain analytical expressions for the spectra providing IP and PZD. \hyperref[Experiment]{The experiment} for the atomic cell with $^{87}$Rb atoms and Ar-N$_2$ buffer gas mixture demonstrates how the nonlinearity affects the error-signal frequency and the way to reduce this influence.

\label{Theory}\section{Theory}
\subsection{Model and initial equations}

\begin{figure}[ht]
\center
{\includegraphics[width=\columnwidth]{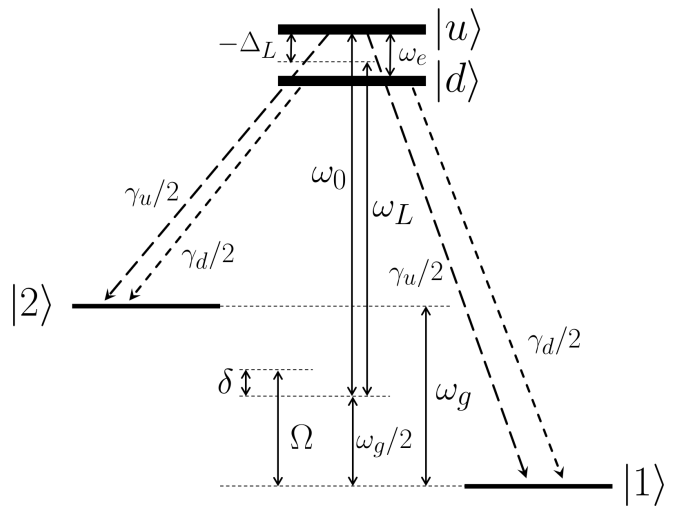}}
\caption{The double $\Lambda$-system of levels under consideration. $\gamma_u$ and $\gamma_d$ are decay rates of the excited state levels $\left|u\right\rangle$ and $\left|d\right\rangle$, respectively. Partial decay rates to the ground states levels are assumed to be the same and equal to $\gamma_u/2$ and $\gamma_d/2$, respectively.}
\label{DoubleLambdaSystem}
\end{figure}

A couple of coherent optical fields tuned to atomic transitions with a common excited state is required for excitation of a CPT resonance~\cite{Alzetta1976,PhysRevA.14.1498,1976NCimL..17..333A,Gray:78}. Considering CPT-based atomic clocks, these transitions are between states ${\mathbf{n}}{{\mathbf{S}}_{{\mathbf{1/2}}}}$ and ${\mathbf{n}}{{\mathbf{P}}_{{\mathbf{1/2}}}}$ of an alkali metal atom, since excitation of a CPT resonance via D$_1$~line provides a higher contrast compared with that via D$_2$~line~\cite{PhysRevA.61.012504,Stahler:02,Lutwak2004}. In compact atomic clocks modulation of the VCSEL injection current with the frequency $\Omega$ is used to obtain a polychromatic optical field for the resonance excitation in a standard scheme with $\sigma$ polarization~\cite{SHAH201021}. The modulation frequency is nearly equal to the half of the ground-state hyperfine splitting $\omega_g/2$, hence for the two-photon detuning $2\delta=2\Omega-\omega_g$ the inequality $\delta\ll\omega_g$ holds. The CPT resonance is excited by the resonant optical field components with frequencies $\omega_L+\Omega$ and $\omega_L-\Omega$, where $\omega_L$ is the carrier frequency tuned to the level $F=I+1/2$ of the excited state ${\mathbf{n}}{{\mathbf{P}}_{{\mathbf{1/2}}}}$. This provides a higher CPT resonance contrast due to a less optical pumping of the sublevel ${\mathbf{n}}{{\mathbf{S}}_{{\mathbf{1/2}}}}\left| {F = I+1/2; m = \pm(I+1/2)} \right\rangle$ compared with the case of coupling via the excited state level $F = I-1/2$~\cite{2017Metro..54..418W}; signs ``$+$'' and ``$-$'' here and after correspond to the $\sigma^+$ and $\sigma^-$ polarizations. For example, in the case of $^{87}$Rb atoms used in the experiment, these transitions are ${\mathbf{5}}{{\mathbf{S}}_{{\mathbf{1/2}}}}\left| {F = 2; m = 0} \right\rangle\rightarrow{\mathbf{5}}{{\mathbf{P}}_{{\mathbf{1/2}}}}\left| {F = 2;m=\pm1} \right\rangle$ and ${\mathbf{5}}{{\mathbf{S}}_{{\mathbf{1/2}}}}\left| {F = 1; m = 0} \right\rangle\rightarrow{\mathbf{5}}{{\mathbf{P}}_{{\mathbf{1/2}}}}\left| {F = 2;m=\pm1} \right\rangle$. If buffer gas is used to decrease the ground-state relaxation rate and (or) to prevent the radiation trapping~\cite{RevModPhys.44.169,vanier1989quantum,SHAH201021,doi:10.1063/1.5026238}, then the optical lines are broadened due to collisions of alkali metal atoms with buffer gas components~\cite{PhysRevA.83.062714,Kroemer:15,PhysRevA.54.2216}. When the broadening makes transitions to $F=I+1/2$ and $F=I-1/2$ levels unresolved, the resonant components are tuned to the maximum of the overall absorption contour. We investigate the case when the absorption lines can overlap, therefore $\omega_L$ is consider as nearly equal to the half-sum of the D$_1$~line transitions to the excited-state level $F=I+1/2$, $\omega_0$. The carrier frequency detuning is defined as $\Delta_L=\omega_L-\omega_0$, which can differ from zero due to an optical field frequency pulling, but by the value much smaller than the excited-state hyperfine frequency splitting, making the inequality $\Delta_L/\omega_e\ll1$ suitable. In the case of overlapping the interaction with ${\mathbf{n}}{{\mathbf{P}}_{{\mathbf{1/2}}}}\left| {F = I-1/2;m=\pm1} \right\rangle$ level gives a significant contribution to the resonance amplitude. The double $\Lambda$-system of levels, see Fig.~\ref{DoubleLambdaSystem}, is considered~\cite{Study} to take into account the effect of the excited-state hyperfine structure on the CPT resonance.

For simplicity the VCSEL's polychromatic spectrum is often considered as corresponding to the following phase-modulated field, $\mathcal{E}(t)=\mathcal{E}\left(e^{-i(\omega_Lt+a\sin{\Omega t})}+c.c.\right)$. It can be written as $\mathcal{E}(t)=\mathcal{E}\left(e^{-i\omega_Lt}\sum_kJ_k(a)e^{-ik\Omega t}+c.c.\right)$, i.e., the amplitude of each spectrum component with index $k$ is determined by the Bessel function of the order $k$. However, a typical VCSEL spectrum under injection current modulation differs from the spectrum of the phase-modulated field: while the frequency spacing between spectrum components $k+1$ and $k$ is equal to $\Omega$, the amplitudes of its spectral components can not be considered as given by the Bessel functions~\cite{doi:10.1063/1.2838175,Gruet:13,al2015vcsels}. Moreover, the VCSEL spectra are asymmetric, i.e., powers of components $|k|$ and $-|k|$ are unequal in the vast majority of cases. On the other hand, amplitudes of the VCSEL spectral components resemble the behavior of dumped oscillations. With the increase of the current modulation depth each of them reaches the first maximum, then passes through a local minimum and after that reaches a new maximum, and so on. Such behavior can provide at least two spectra for which IP can be observed~\cite{BGpaper}. The most significant factor of the VCSEL spectra asymmetry is the inequality of the resonant components amplitudes ($k=\pm1$), since this imbalance decreases contrast of CPT resonance, makes it asymmetric and gives additional contribution to the shift~\cite{Stahler:02,doi:10.1063/1.5026238,4623084}.

Due to the mentioned above features of the VCSELs spectra, we consider the following polychromatic optical field:
\begin{equation}
\begin{gathered}
\mathcal{E}(t)=\frac12e^{-i\omega_Lt}\bigg(\sum_{k\neq\mp1}\mathcal{E}_ke^{-ik\left(\Omega t+\varphi(t)\right)}-\\
-\mathcal{E}_{-1}e^{i\left(\Omega t+\varphi(t)\right)}+\mathcal{E}_1e^{-i\left(\Omega t+\varphi(t)\right)}\bigg)+c.c.,
\end{gathered}
\label{OpticalField}
\end{equation}

\noindent where amplitudes are assumed to be real, $\mathcal{E}_k=\mathcal{E}^*_k$, and positive. While Eq.~\eqref{OpticalField} is formally written with the infinite number of components, the operating spectra of VCSELs providing the zero light shift in practice have only a few higher sidebands~\cite{BGpaper}.

We consider modulation spectroscopy of the CPT resonances~\cite{1402-4896-93-11-114002}. The crucial point of this technique is the time-harmonic perturbation of the $\Omega$ frequency's phase with the index $a$, $\varphi(t)=a\sin{\omega_mt}$. This modulation provides oscillating response at the frequency $\omega_m$ in the absorption coefficient of the optical field by an atomic ensemble $\kappa(t)$. Then the synchronous detection is used to obtain dispersive-shape response for stabilization of the LO frequency and tracking of the CPT resonance frequency. We note that $\omega_m$ is comparable to the CPT resonance width, which ensures the fulfillment of the inequality $\omega_m\ll\omega_g,\Omega$. The typical CPT resonance width is in the range from a few tenths to several kHz, depending on atomic cell size and intensity of an optical field, while the ground-state hyperfine splitting of the alkali metal atoms is in the GHz range. Also, to avoid confusion, we define the in-phase and quadrature signals with respect to the modulation of the frequency $\Omega$, 
not to modulation of its phase, i.e., amplitudes $S$ and $Q$ of the in-phase and quadrature signals, read as
\begin{subequations}
\begin{equation}
S=\int dt\,\kappa(t)\cos{\omega_mt},
\label{Sdefinition}
\end{equation}
\begin{equation}
Q=\int dt\,\kappa(t)\sin{\omega_mt},
\label{Qdefinition}
\end{equation}
\label{SandQdefinition}
\end{subequations}

\noindent respectively, where integration is carried out over a few periods of the frequency $\omega_m$, and $\kappa(t)$ is the absorption coefficient. The synchronous detection can be also made with some phase shift, i.e., when integrands in Eqs.~\eqref{Sdefinition}-\eqref{Qdefinition} contain $\cos{(\omega_mt+\alpha)}$ and $\sin{(\omega_mt+\alpha)}$, which gives some combinations of the in-phase and quadrature signals. These combinations can have a steeper slope that $S$ or $Q$ slopes and can provide higher feedback level for stabilization of the LO frequency.

Since the CPT effect is mainly induced by the resonant components, the following notations are used further for convenience: $V_{L_u,L_d}=d_{u,d}\mathcal{E}_{-1}/2\hbar$, $V_{R_u,R_d}=d_{u,d}\mathcal{E}_1/2\hbar$, and $V_{k_u,k_d}=d_{u,d}\mathcal{E}_k/2\hbar$, where $d_u$ and $d_d$ are defined as $d_u=\left\langle u\right|\hat{d}\left|2\right\rangle=\left\langle u\right|\hat{d}\left|1\right\rangle$ and $d_d=\left\langle d\right|\hat{d}\left|2\right\rangle=\left\langle d\right|\hat{d}\left|1\right\rangle$, respectively, and are assumed to be real. As one can see, the Rabi frequencies $V_{L_u,L_d}$ and $V_{R_u,R_d}$ are associated with optical transitions for ``left shoulder'' (low-frequency ones $|2\rangle\rightarrow|u\rangle,|d\rangle$) and ``right shoulder'' (high-frequency ones $|1\rangle\rightarrow|u\rangle,|d\rangle$) of the double $\Lambda$-system of levels under consideration, respectively. For completeness, we note that $d_u/d_d=\sqrt{3}$ for the $^{87}$Rb atoms due to the different values of the Clebsch-Gordan coefficients. The values of other parameters will be also estimated for the $^{87}$Rb atoms.

So far as buffer gas is used to reduce relaxation rate of the microwave coherence arising due to collisions of alkali metal atoms with the atomic cell walls and to quench their fluorescence~\cite{RevModPhys.44.169,vanier1989quantum,SHAH201021,doi:10.1063/1.5026238}, the phenomenological constant $\Gamma$ is used to describe resulting broadening of the optical absorption line. The relaxation of the ground-state elements is assumed to be isotropic: one constant $\Gamma_g$ is used to describe decoherence of the dark state and relaxation of the ground-state populations to their equilibrium values $1/2$. Operating pressure of buffer gas in atomic cells is several tens of Torr, depending on atomic cell size~\cite{vanier1989quantum,doi:10.1063/1.1494115,Knappe:04}, therefore we neglect the Doppler broadening since it is compared to or smaller than the homogeneous width of the optical line. On the opposite, we neglect the microwave Doppler effect since it is canceled due to the Dicke narrowing~\cite{PhysRev.89.472}. In addition to the mentioned above assumptions, the strong inequality $(\Gamma/\Omega)^2\ll1$ holds at such buffer gas pressures.

The described above approach can be summarized in the following Hamiltonian $\hat{\mathcal{H}}(t)$ and the relaxation term $\hat{\Gamma}\hat{\rho}(t)$,
\begin{equation*}
\hat{\mathcal{H}}(t)=\begin{pmatrix}
\hbar\omega_0 & 0 & -d_u\mathcal{E}(t) & -d_u\mathcal{E}(t) \\
0 & \hbar(\omega_0-\omega_e) & -d_d\mathcal{E}(t) & -d_d\mathcal{E}(t) \\
-d_u\mathcal{E}(t) & -d_d\mathcal{E}(t) & \hbar\omega_g/2 & 0 \\ 
-d_u\mathcal{E}(t) & -d_d\mathcal{E}(t) & 0 & -\hbar\omega_g/2
\end{pmatrix},
\end{equation*}
{\small
\begin{equation*}
\begin{aligned}
&\hat{\Gamma}\hat{\rho}(t)\\
&=\begin{pmatrix}
\gamma_u\rho_{uu} & 0 & \Gamma\rho_{u2} & \Gamma\rho_{u1} \\
0 & \gamma_d\rho_{dd} & \Gamma\rho_{d2} & \Gamma\rho_{d1} \\
\Gamma\rho_{2u} & \Gamma\rho_{d2} & \begin{array}{c}\Gamma_g\left(\rho_{22}-\frac12\right) \\ -\frac{\gamma_u}{2}\rho_{uu}-\frac{\gamma_d}{2}\rho_{dd} \end{array} & \Gamma_g\rho_{21} \\
\Gamma\rho_{1u} & \Gamma\rho_{d1} & \Gamma_g\rho_{12} & \begin{array}{c}\Gamma_g\left(\rho_{11}-\frac12\right)\\ -\frac{\gamma_u}{2}\rho_{uu}-\frac{\gamma_d}{2}\rho_{dd} \end{array} \\
\end{pmatrix},
\end{aligned}
\end{equation*}}
\noindent which obey the von Neumann equation
\begin{equation}
i\hbar\left(\dfrac{\partial}{\partial t}+\hat{\Gamma}\right)\hat{\rho}(t)=[\hat{\mathcal{H}}(t),\hat{\rho}(t)].
\label{vonNeumannEquation}
\end{equation}

\subsection{Analytical solution}
\label{Analytics}

We neglect excited-state populations $\rho^{ee}_{uu}$, $\rho^{ee}_{dd}$ compared with ground-state ones, $\rho^{gg}_{22}$ and $\rho^{gg}_{11}$, under assumption of the low saturation regime, $V^2_{L_u,L_d}/(\gamma_{u,d}\Gamma)\ll1$, $V^2_{R_u,R_d}/(\gamma_{u,d}\Gamma)\ll1$, which holds with a wide margin for the case of CPT-based clocks. The coherence $\rho^{ee}_{ud}$ can be also neglected since $\omega_e$ is smaller by almost an order of magnitude compared with $\omega_g$, i.e., the optical field components do not induce any V-resonances. Finally, the standard approach --- the resonant approximation for the optical field: $\left\{\rho^{eg}_{u2},\rho^{eg}_{d2}\right\}=\left\{\rho_{u2},\rho_{d2}\right\}e^{-i(\omega_L-\Omega)}$; $\left\{\rho^{eg}_{u1},\rho^{eg}_{d1}\right\}=\left\{\rho_{u1},\rho_{d1}\right\}e^{-i(\omega_L+\Omega)}$; $\rho^{gg}_{21}=\rho_{21}e^{-2i\Omega t}$, and the adiabatic elimination of the excited state (which means that obtained solution is correct at times $t\gg1/\gamma_{u,d},1/\Gamma$), allows us to express the excited-state populations via density matrix elements of the ground state:
\begin{subequations}
\begin{equation*}
\rho_{uu}=\dfrac{2\Gamma\left(V^2_{L_u}\rho_{22}+V^2_{R_u}\rho_{11}-2V_{L_u}V_{R_u}\text{Re}\left\{\rho_{21}\right\}\right)}{\gamma_u(\Delta^2_L+\Gamma^2)},
\end{equation*}
\begin{equation*}
\rho_{dd}=\dfrac{2\Gamma\left(V^2_{L_d}\rho_{22}+V^2_{R_d}\rho_{11}-2V_{L_d}V_{R_d}\text{Re}\left\{\rho_{21}\right\}\right)}{\gamma_d\left[(\Delta_L+\omega_e)^2+\Gamma^2\right]},
\end{equation*}
\end{subequations}

\noindent demonstrating that for the case of moderate buffer gas pressures, when $(\Gamma/\omega_e)^2\ll1$ (but $\Gamma\lesssim\omega_e$), population $\rho_{dd}$ is $\Gamma/\omega_e$ times smaller than $\rho_{uu}$ population. Note that we have omitted the superscripts for brevity. We remind that $\gamma_{u,d}\propto d^2_{u,d}$, therefore the total population of the excited state is simply given by
\begin{equation}
\rho_{uu}+\rho_{dd}=2\dfrac{P}{\gamma\Gamma}\big(\mathcal{V}^2_L\rho_{22}+\mathcal{V}^2_R\rho_{11}-2\mathcal{V}_L\mathcal{V}_RRe\{\rho_{21}\}\big),
\label{eepopulation}
\end{equation}
\noindent where $P=\Gamma^2/(\Delta^2_L+\Gamma^2)+\Gamma^2/[(\Delta_L+\omega_e)^2+\Gamma^2]$, $\mathcal{V}_{L,R}=d\mathcal{E}_{-1,1}/(2\hbar)$, $d$ is the reduced dipole matrix element.

An important note must be made regarding Eq.~\eqref{eepopulation}. Only contribution of the resonant optical field components were explicitly accounted for its derivation, which is correct for $(\Gamma/\Omega)^2\ll1$. This feature has obvious physical meaning: while the optical line broadening is much smaller than the frequency spacing between optical field components, the CPT resonance is induced mainly by the first sidebands of the optical field. Although frequencies of the components $k+1$ and $k-1$ ($k\neq0$) satisfy to the condition of two-photon resonance, they have large detunings from the optical transition and their explicit contribution to the CPT resonance can be neglected. If the last strong inequality does not hold, a contribution of other pairs of components should be accounted, for example, $\propto\mathcal{E}_{-2}\mathcal{E}_0$, $\propto\mathcal{E}_0\mathcal{E}_2$. For the case $(\Gamma/\Omega)^2\ll1$ under consideration in this paper, the non-resonant components of the optical field contribute only to the light shift, see the Eq.~\eqref{rho21} and the paragraph after it.

%As one can also notice,
The modulation term $a\sin{\omega_m t}$ is not contained explicitly in the populations of the excited state. This is due to the fact that $\omega_m\ll\gamma_{u,d},\Gamma$ and the modulation index $a$ is moderate, $a\lesssim1$, i.e., the time-perturbation of $\Omega$ is too ``weak'' and slow compared to the excited-state relaxation processes to directly affect its populations and coherencies. The situation is opposite for the ground-state density-matrix elements, for which we obtain the following system of equations under the mentioned above assumptions:
\begin{subequations}
{\small
\begin{equation}
\frac{\partial}{\partial t}\rho_{22}=-V_L\rho_{22}+V_R\rho_{11}-i\mathcal{K}\left(\rho_{21}-\rho_{12}\right)-\Gamma_g(\rho_{22}-\frac12),
\label{22population}
\end{equation}
\begin{equation}
\frac{\partial}{\partial t}\rho_{11}=-V_R\rho_{11}+V_L\rho_{22}+i\mathcal{K}\left(\rho_{21}-\rho_{12}\right)-\Gamma_g(\rho_{11}-\frac12),
\label{11population}
\end{equation}
\begin{equation}
\begin{gathered}
\left[i\frac{\partial}{\partial t}+2\left(\tilde{\delta}+a\omega_m\cos{\omega_m t}\right)+i\tilde{\Gamma}_g\right]\rho_{21}=\\
=iV_{LR}+\mathcal{K}\left(\rho_{22}-\rho_{11}\right),
\end{gathered}
\label{rho21}
\end{equation}}
\label{initialequations}
\end{subequations}

\noindent where the following notations were introduced: $V_L=V^2_{L_u}\frac{\Gamma}{\Delta^2_L+\Gamma^2}+V^2_{L_d}\frac{\Gamma}{(\Delta_L+\omega_e)^2+\Gamma^2}$, $V_R=V^2_{R_u}\frac{\Gamma}{\Delta^2_L+\Gamma^2}+V^2_{R_d}\frac{\Gamma}{(\Delta_L+\omega_e)^2+\Gamma^2}$, $V_{LR}=\Gamma\left(\frac{V_{L_u}V_{R_u}}{\Delta^2_L+\Gamma^2}+\frac{V_{L_d}V_{R_d}}{(\Delta_L+\omega_e)^2+\Gamma^2}\right)$, $\mathcal{K}=\Delta_L\frac{V_{L_u}V_{R_u}}{\Delta^2_L+\Gamma^2}+\left(\Delta_L+\omega_e\right)\frac{V_{L_d}V_{R_d}}{(\Delta_L+\omega_e)^2+\Gamma^2}$. The relaxation term $\tilde{\Gamma}_g$ of the coherence takes into account the power broadening, $\tilde{\Gamma}_g=\Gamma_g+V_L+V_R$. The two-photon detuning $\tilde{\delta}$ accounts for the light shift, $2\tilde{\delta}=2\delta+\delta_r+\delta_{nr}$, where $\delta_r$ is the resonant light shift, and $\delta_{nr}$ is the non-resonant light shift~\cite{refId0,PhysRevA.58.2345,827437,1402-4896-93-11-114002} from all components of the optical field~\eqref{OpticalField}, $\delta_r=-\Delta_L\frac{V^2_{L_u}-V^2_{R_u}}{\Delta^2_L+\Gamma^2}-\left(\Delta_L+\omega_e\right)\frac{V^2_{L_d}-V^2_{R_d}}{(\Delta_L+\omega_e)^2+\Gamma^2}$, $\delta_{nr}=-\sum_{k\neq-1}\frac{V^2_{-k_u}}{(k+1)\Omega}+\sum_{k\neq1}\frac{V^2_{-k_u}}{(k-1)\Omega}$ $
-\sum_{k\neq-1}\frac{V^2_{-k_d}}{(k+1)\Omega}+\sum_{k\neq1}\frac{V^2_{-k_d}}{(k-1)\Omega}$.

{\setstretch{0.95}
The simplest analytical solution of Eqs.~\eqref{22population}-\eqref{rho21} can be obtained in the case $a\ll1$ via the Fourier series expansion of the density matrix elements (which means that the solution is quasi-stationary, which is valid at $t\gg1/\tilde{\Gamma}_g$), $\rho_{22}=\sum_kG_ke^{-ik\omega_mt}$ (wherein $G_k=G^*_{-k}$) and $\rho_{21}=\sum_kC_ke^{-ik\omega_mt}$. This gives the following set of equations with holding the terms over $a^2$ and taking into account that $C_{k,-k}\propto J_0(a)J_k(a)\propto a^k$, $G_k\propto J_0(a)J_k(a)\propto a^k$~\cite{10.1088/1742-6596/941/1/012055}:
\begin{subequations}
\begin{equation}
(2\tilde{\delta}+i\tilde{\Gamma}_g)C_0+a\omega_m(C_1+C_{-1})=iV_{LR}+\mathcal{K}(2G_0-1),
\end{equation}
\begin{equation}
(2\tilde{\delta}+\omega_m+i\tilde{\Gamma}_g)C_1+a\omega_mC_0=2\mathcal{K}G_1,
\end{equation}
\begin{equation}
(2\tilde{\delta}-\omega_m+i\tilde{\Gamma}_g)C_{-1}+a\omega_mC_0=2\mathcal{K}G^*_1,
\end{equation}
\begin{equation}
(2\tilde{\delta}+2\omega_m+i\tilde{\Gamma}_g)C_2+a\omega_mC_1=2\mathcal{K}G_2,
\end{equation}
\begin{equation}
(2\tilde{\delta}-2\omega_m+i\tilde{\Gamma}_g)C_{-2}+a\omega_mC_{-1}=2\mathcal{K}G^*_2,
\end{equation}
\begin{equation}
i\tilde{\Gamma}_gG_0=\mathcal{K}(C_0-C^*_0)+i(V_R+\Gamma_g/2),
\end{equation}
\begin{equation}
(\omega_m+i\tilde{\Gamma}_g)G_1=\mathcal{K}(C_1-C^*_{-1}),
\end{equation}
\begin{equation}
(2\omega_m+i\tilde{\Gamma}_g)G_2=\mathcal{K}(C_2-C^*_{-2}),
\end{equation}
\label{smallindexsystem}
\end{subequations}

\noindent while, as follows from Eqs.~\eqref{Sdefinition}-\eqref{Qdefinition} and Eq.~\eqref{eepopulation}, amplitudes $S$ and $Q$ are expressed as
\begin{subequations}
\begin{equation}
S\propto\left(\mathcal{V}^2_L-\mathcal{V}^2_R\right)\text{Re}\left\{G_1\right\}-\mathcal{V}_L\mathcal{V}_R\text{Re}\left\{C_1+C_{-1}\right\},
\end{equation}
\begin{equation}
Q\propto\left(\mathcal{V}^2_L-\mathcal{V}^2_R\right)\text{Im}\left\{G_1\right\}-\mathcal{V}_L\mathcal{V}_R\text{Im}\left\{C_1-C_{-1}\right\}.
\end{equation}
\label{SandQviaFourierAmplitudes}
\end{subequations}

For simplicity we derive further expressions under strong but appropriate for the experimental case inequalities $(\Delta_L/\Gamma)^2\ll1$, $(d_d/d_u)^2(\Gamma/\omega_e)^2\ll1$, which allow us to neglect $\mathcal{K}^2$ compared with $\tilde{\Gamma}^2_g$. We note that holding the terms over $a^2$ provides reasonable accuracy of Eqs.~\eqref{smallindexsystem} for $a\lesssim1/2$, which follows from properties of the Bessel functions and the fact that the responses have the cubic correction over $a$.}

\subsection{Sources of the effect}
\subsubsection{Inequality of the resonant components powers and high values of the modulation frequency}
\label{SubsectionAsymmetry}

By linearizing the solution of Eqs.~\eqref{smallindexsystem} over $\tilde{\delta}$ we obtain the following expressions for amplitudes of the in-phase and quadrature signals:
{\footnotesize
\begin{subequations}
\begin{equation}
S=\frac{2P}{\gamma\Gamma}\frac{4a\omega_mV_{LR}\left[2\tilde{\delta}\mathcal{V}_L\mathcal{V}_R\tilde{\Gamma}^2_g+\mathcal{K}(\mathcal{V}^2_L-\mathcal{V}^2_R)(\tilde{\Gamma}^2_g+\omega^2_m)\right]}{\tilde{\Gamma}_g\left(\tilde{\Gamma}^2_g+\omega^2_m\right)^2},
\label{SSmall}
\end{equation}
\begin{equation}
Q=\frac{2P}{\gamma\Gamma}\frac{4a\omega^2_mV_{LR}\left[\tilde{\delta}\mathcal{V}_L\mathcal{V}_R[3\tilde{\Gamma}^2_g+\omega^2_m]+\mathcal{K}(\mathcal{V}^2_L-\mathcal{V}^2_R)(\tilde{\Gamma}^2_g+\omega^2_m)\right]}{\tilde{\Gamma}^2_g\left(\tilde{\Gamma}^2_g+\omega^2_m\right)^2}.
\label{QSmall}
\end{equation}
\end{subequations}}

As one can see from the terms in braces in Eq.~\eqref{SSmall} and Eq.~\eqref{QSmall}, the zero-crossing points have additional shift which differs from non-resonant and resonant light shifts. In particular, we have for the frequency shift of the in-phase response's zero:
\begin{subequations}
\begin{equation}
S= A(2\delta+\delta_r+\delta_{nr}+\delta_{as}),
\label{InPhPZD}
\end{equation}
\begin{equation}
A=\dfrac{2P}{\gamma\Gamma}\,\dfrac{4a\omega_mV_{LR}\mathcal{V}_L\mathcal{V}_R\tilde{\Gamma}_g}{\left(\tilde{\Gamma}^2_g+\omega^2_m\right)^2},
\end{equation}
\begin{equation}
\delta_{as}=\mathcal{K}\dfrac{\mathcal{V}^2_L-\mathcal{V}^2_R}{\mathcal{V}_L\mathcal{V}_R}\dfrac{\tilde{\Gamma}^2_g+\omega^2_m}{\tilde{\Gamma}^2_g}.
\end{equation}
\label{SShift}
\end{subequations}

There are two terms, $\delta_r$ and $\delta_{as}$, emerging at $\mathcal{E}_{-1}\neq\mathcal{E}_1$. While the resonant light shift is a consequence of the changed energy of the ground-state levels,
the additional shift $\delta_{as}$ arises due to asymmetry of the in-phase signal (it is not an odd function over $\tilde{\delta}$). The term $\delta_{as}$ leads to the following features of the in-phase signal zero-crossing point. Firstly, while PZDs are determined by equality of the term in parentheses of~\eqref{InPhPZD} to zero, the condition on IPs reads as
\begin{equation}
\dfrac{\partial}{\partial \mathcal{E}^2}\bigg(\delta_{r}+\delta_{nr}+\delta_{as}\bigg)=0,
\end{equation}

\noindent where for convenience we have introduced the notation $\mathcal{E}^2=\sum_k\mathcal{E}^2_k$ and $\mathcal{E}^2_k=\sigma_k\mathcal{E}^2$, $\sum_k\sigma_k=1$. The term $\delta_r+\delta_{nr}+\delta_{as}$ and its partial derivative over $\mathcal{E}^2$ turn to zero for different sets of $\sigma_k$ (for different spectra) since it is a nonlinear function of $\mathcal{E}^2$. It is because $\delta_{as}$ is not proportional to $\mathcal{E}^2$ (due to the term $\propto\omega^2_m$ if the power broadening is compared with the ground-state relaxation), while $\delta_r\propto\mathcal{E}^2$ and $\delta_{nr}\propto\mathcal{E}^2$. Therefore at IPs, the in-phase response crosses zero at $\delta_0\neq0$ given by $2\delta_0=-\delta_r-\delta_{nr}-\delta_{as}$. Secondly, the insensitivity and the zero displacement can be obtained simultaneously only at $(\omega_m/\tilde{\Gamma}_g)^2\ll1$. More important is that in this case the suppression of the sensitivity is total, while for $(\omega_m/\tilde{\Gamma}_g)^2\sim1$ only the linear response of the error-signal frequency to variations in $\mathcal{E}^2$ vanishes.

\begin{figure}[!hb]
\begin{minipage}{\textwidth}
{\centering
\includegraphics[width=0.65\columnwidth]{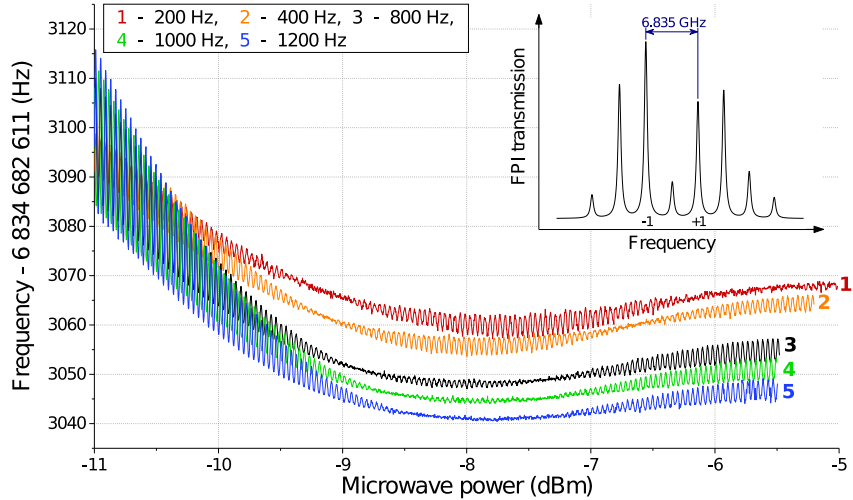}
\caption{Dependencies of the CPT resonance frequency on the depth of the injection current modulation (RF power) taken with simultaneous modulation of the light intensity obtained for different modulation frequencies at $a=1/2$. The atomic cell temperature is $45$~$^{\circ}$C. The inset shows the spectrum corresponding to the first IP at $\omega_m=200$~Hz.}
\label{Exp1}}
\end{minipage}
\end{figure}

Since $\delta_{as}$ contains the term $\propto\omega^2_m$, one can expect to observe different values of $\delta_0$ for IPs at values of $\omega_m$ compared with the CPT resonance width, see Fig.~\ref{Exp1} and Table~\ref{Table1}. On the other hand, different frequencies of the error signal at IPs are the unambiguous criterion that they are shifted from the frequency of the ground-state transition unperturbed by optical fields.
\onecolumngrid
\clearpage
\twocolumngrid

\subsubsection{Optically thick medium}

In practice, the necessity to have a higher contrast of the CPT resonance implies an increase of the atomic cell temperature. In such a situation absorption of the resonant sidebands should be accounted. Two regimes of the absorption can be distinguished in the model of $\Lambda$-system of levels in the case of neglecting the Doppler broadening and $\Delta_L=0$, $V_L=V_R=V$~\cite{Agapev:1993,Godone2002}. At the first one the contrast is low and the resonant sidebands decay exponentially over the atomic cell length. It is determined by the condition $V\ll\Gamma_g$ for the  light incident on an atomic cell. In the second one, provided by the opposite inequality, the contrast is high, and the first sidebands decay linearly. At first glance the second regime looks attractive to raise the CPT resonance contrast by the increase of optical field intensity. But in practice, the opposite situation occurs due to the optical pumping of alkali-metal atoms to the end sublevel ${\mathbf{n}}{{\mathbf{S}}_{{\mathbf{1/2}}}}\left|{F=I+1/2;m_F=I+1/2} \right\rangle$, therefore the maximal contrast is obtained in the case $V\sim\Gamma_g$~\cite{PhysRevA.67.065801}. The first sidebands are also absorbed on the magnetic sublevels, for which condition of the two-photon resonance is not fulfilled, i.e., the first regime of the absorption takes place. Therefore the absorption of the resonant optical field components can be accounted phenomenologically by introducing the decay as $\mathcal{E}^2_{-1,1}\rightarrow\mathcal{E}^2_{-1,1}e^{-\beta z}$. Then the frequency $\delta_0$ of the zero-crossing point of the in-phase signal is determined by the following condition (here we consider the symmetrical spectra):
\begin{subequations}
\begin{equation}
2\delta_0=-\dfrac{\int^l_0A(z)\delta_{nr}(z)dz}{\int^l_0A(z)dz},
\end{equation}
\begin{equation}
A(z)=\dfrac{2P}{\gamma\Gamma}\dfrac{4a\omega_m\bar{V}^2(z)\mathcal{V}^2(z)\tilde{\Gamma}_g(z)}{\left[\tilde{\Gamma}^2_g(z)+\omega^2_m\right]^2},
\end{equation}
\label{ThickMediumFrquency}
\end{subequations}

\noindent where $\bar{V}^2(z)=\Gamma\left[\frac{V^2_u}{\Delta^2_L+\Gamma^2}+\frac{V^2_d}{(\Delta_L+\omega_e)^2+\Gamma^2}\right]e^{-\beta z}$, $\mathcal{V}^2(z)=\mathcal{V}^2e^{-\beta z}$, $\tilde{\Gamma}_g(z)=\Gamma_g+2V^2e^{-\beta z}/\Gamma$. Integration in Eqs.~\eqref{ThickMediumFrquency} means averaging of the in-phase signal over the length of an atomic cell. The expression demonstrates that PZDs are determined by $\int^l_0A(z)\delta_{nr}(z)dz=0$.

We again use the parameters $\mathcal{E}^2=\sum_k\mathcal{E}^2_k$ and $\sigma_k=\mathcal{E}^2_k/\mathcal{E}^2$ and write the condition on IPs as
\begin{equation}
\begin{gathered}
\int^l_0\frac{\partial A(z)\delta_{nr}(z)}{\partial\mathcal{E}^2}dz\int^l_0A(z)dz\\
-\int^l_0\frac{\partial A(z)}{\partial\mathcal{E}^2}dz\int^l_0A(z)\delta_{nr}(z)dz=0.
\end{gathered}
\label{condition}
\end{equation}

\begin{figure}[!hb]
\begin{minipage}{\textwidth}
{\centering
\includegraphics[width=0.7\columnwidth]{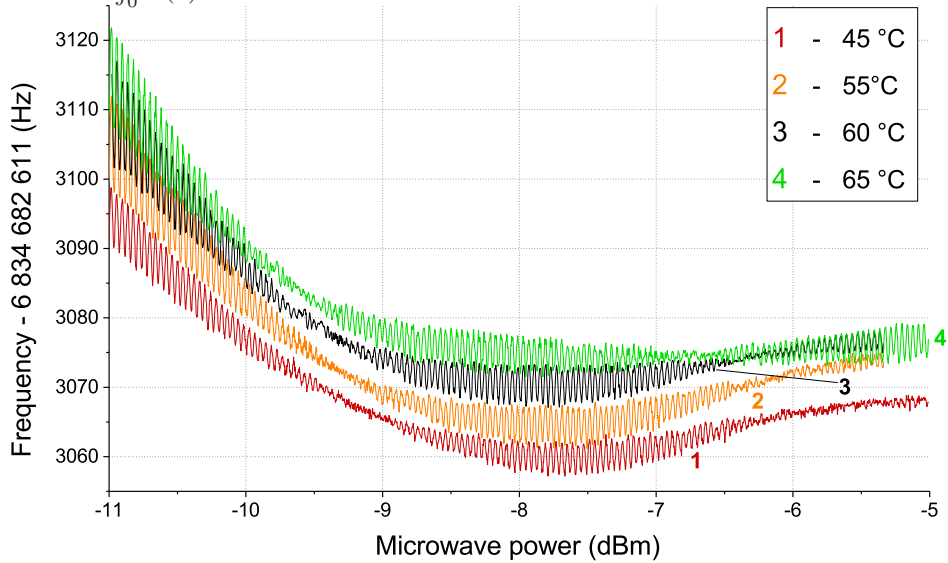}
\caption{(a) Dependencies of the CPT resonance frequency on the RF power taken with simultaneous modulation of the light intensity at different temperatures. The modulation frequency is $200$~Hz and $a=1/2$.}
\label{Exp2}}
\end{minipage}
\end{figure}

It demonstrates the same situation as was mentioned in section~\ref{SubsectionAsymmetry}: the frequency shift $\delta_0$ is a nonlinear function of $\mathcal{E}^2$, therefore properly chosen spectra can provide only vanishing of the linear response, not the total insensitivity, and when condition~\eqref{condition} is fulfilled, then $\delta_0\neq0$. And vice versa, at PZDs, the error-signal frequency remains sensitive to variations in the intensity. As a consequence, one can observe different frequencies of the zero-crossing point at IPs with growth of the absorption level, even if they coincide in the optically thin media, see Fig.~\ref{Exp2} and Table~\ref{Table2}. Also, as in the case of optically thin medium and $\mathcal{E}_{-1}\neq\mathcal{E}$, the difference of frequencies means that they are shifted from the frequency of ``$0-0$'' transition unperturbed by optical fields.
\onecolumngrid
\clearpage
\twocolumngrid

\label{Experiment}\section{Experiment}

\begin{figure}[ht]
\centering 
\includegraphics[width=\columnwidth]{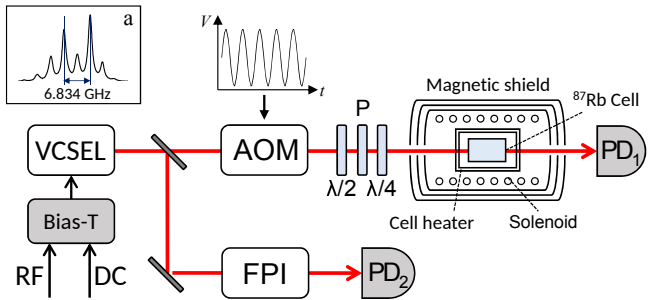}
\caption{The layout of experimental setup. VCSEL is a vertical-cavity surface-emitting laser, whose frequency is tuned to $^{87}$Rb D$_1$~line. Bias-T combines injection DC with AC of the microwave modulation (RF). AOM is an acousto-optic modulator, FPI is a Fabry-Perot interferometer, PD$_{1,2}$ are photodetectors, $\lambda/2$ is a half-wave plate, $\lambda/4$ is a quarter-wave plate, P is a polarizer.
}
\label{Exp_scheme}
\end{figure}

The experimental setup is schematically shown in Fig.~\ref{Exp_scheme}. A single-mode $795$~nm VCSEL is current-modulated at the frequency $\Omega\approx3.417$~GHz by a microwave synthesizer. The first-order sidebands of the polychromatic spectrum (see the inset in Fig.~\ref{Exp1}) excite the CPT resonance in a standard scheme using a circular polarization~\cite{SHAH201021,doi:10.1063/1.5026238}. Coarse laser intensity control is performed by a half-wave plate and a polarizer. A longitudinal magnetic field ($\sim0.1$~G) produced by a solenoid is used to resolve the metrological and two magneto-sensitive resonances. The optical signal is registered by the photodetector PD$_1$ and is used to stabilize both the laser and the LO frequencies. The laser frequency is stabilized by a servo-loop controlling the laser temperature. The LO frequency is stabilized to the zero-crossing point frequency of the in-phase response. The laser spectrum is registered by a scanning Fabry-Perot interferometer with a free spectral range of about $35$~GHz and the photodetector PD$_2$.

The atomic cell of the cylindrical geometry ($8$~mm in diameter and $15$~mm in length) filled with isotopically enriched $^{87}$Rb and a mixture of Ar and N$_2$ is used. The buffer gas pressure is $19$~Torr and the ratio of the partial pressures, P$_{Ar}$/P$_{N_2}$, is~$1.3$. A heater and a servo-loop maintain the temperature with an accuracy of $0.01$~$^{\circ}$C. The cell, heater, and solenoid are placed inside a three-layer magnetic shield isolating the cell from external magnetic fields. The laser beam diameter inside the cell is about $3$~mm, the laser radiation power entering the atomic cell is $62.5$~$\mu$W. The power broadening is $\sim1.2$~kHz and the full width at half maximum of the CPT resonance is $1.6$~kHz (in the $2\delta$ scale) at the RF power $-9$~dBm.

We use the following technique to find RF modulation powers at which the VCSEL's spectrum provides IPs. The light intensity is harmonically varied by an acousto-optic modulator while the RF power is slowly increased. At certain values of the modulation depth (RF power) the frequency response vanishes, i.e., changes of the in-phase signal frequency due to the light intensity modulation are indistinguishable from fluctuations of the frequency caused by other factors. These values are IPs by definition.

The following experimental data are obtained by the mentioned above method. The RF power is scanned from $-11$~dBm to $-5$~dBm during $3900$~s, while the light modulation period is $25$~s (depth of the intensity modulation is~$30$\%). In this case, the servo-loops properly respond to changes in the RF power and the light intensity, and the CPT resonance frequency adiabatically follows changes in the optical field intensity and the RF power.

The dependencies obtained for $a=1/2$ and different values of $\omega_m$ at $45$~$^{\circ}$C are shown in Fig.~\ref{Exp1}; for different temperatures at $\omega_m/2\pi=200$~Hz are shown in Fig.~\ref{Exp2}. As $\omega_m$ grows, IPs move towards each other in the horizontal direction and down in the vertical direction. As the absorption rate increases, IPs move to the left in the horizontal direction and up in the vertical direction. The frequencies corresponding to the first ($\omega_F$) and second ($\omega_S$) IP are presented in Tables~\ref{Table1}-\ref{Table2}. The values of errors were calculated based on the range without distinct frequency fluctuations.
\vfill
\begin{table}[!hb]
\caption{Frequencies of IPs obtained at $45$~$^{\circ}$C for different values of the modulation frequency~$\omega_m$.}
\begin{tabularx}{\columnwidth}{ 
  | >{\centering\arraybackslash}X 
  | >{\centering\arraybackslash}X 
  | >{\centering\arraybackslash}X | } 
 \hline
 $\omega_m/2\pi$,~Hz & $\omega_F/2\pi$\textbackslash error,~Hz & $\omega_S/2\pi$\textbackslash error,~Hz \\
 \hline
 200 & 3067\textbackslash1 & 3067\textbackslash1 \\
 \hline
 400 & 3061.5\textbackslash0.3 & 3059.5\textbackslash0.5 \\
 \hline
 800 & 3051\textbackslash0.3 & 3048.95\textbackslash0.15 \\
 \hline
 1000 & 3046\textbackslash0.2 & 3045.2\textbackslash0.2 \\
 \hline
 1200 & 3041.6\textbackslash$-$ & 3041.2\textbackslash$-$ \\
 \hline
\end{tabularx}
\label{Table1}
\end{table}
\vfill
\begin{table}[!ht]
\caption{Frequencies of IPs obtained at $\omega_m/2\pi=200$~Hz for different temperatures.}
\begin{tabularx}{\columnwidth}{ 
  | >{\centering\arraybackslash}X 
  | >{\centering\arraybackslash}X 
  | >{\centering\arraybackslash}X | } 
 \hline
 Temperature,~$^{\circ}$C & $\omega_F/2\pi$\textbackslash error,~Hz & $\omega_S/2\pi$\textbackslash error,~Hz \\
 \hline
 45 & 3067\textbackslash1 & 3067\textbackslash1 \\
 \hline
 55 & 3072\textbackslash1 & 3071.2\textbackslash0.8 \\
 \hline
 60 & 3076.8\textbackslash0.8 & 3074.4\textbackslash0.4 \\
 \hline
 65 & 3084\textbackslash1 & 3074.5\textbackslash0.5 \\
 \hline
\end{tabularx}
\label{Table2}
\end{table}

Considering dependencies presented in Fig.~\ref{Exp1}, the homogeneous broadening $\Gamma/2\pi$ and the one-photon detuning $\Delta_L/2\pi$, obtained by fitting the optical line, is nearly $380$~MHz and $-28$~MHz, respectively. At such values the parameter $\mathcal{K}$ is positive, and since $\mathcal{E}^2_{-1}$ is greater than $\mathcal{E}^2_1$ (in the entire range of the RF power from $-11$ to $-5$~dBm), the term $(\mathcal{V}^2_L-\mathcal{V}^2_R)/(\mathcal{V}_L\mathcal{V}_R)\partial\mathcal{K}(\omega_m/\tilde{\Gamma}_g)^2/\partial\mathcal{E}^2$ increases the partial derivative over $\mathcal{E}^2$ of the non-resonant light shift induced by the carrier and first sidebands (we remind that the power broadening $V_L+V_R$ is greater than $\Gamma_g$). Therefore, as $\omega_m$ grows, the greater (smaller) RF power are required to obtain the first (second) IP. It appears that under the experimental conditions the difference of $\omega_{F,S}/2\pi$ from frequencies of PZDs is of the order of $30$~Hz for $\omega_m/2\pi=1200$~Hz, if we neglect such a difference for $\omega_m/2\pi=200$~Hz. Moreover, the term $\partial\delta_{as}/\partial\mathcal{E}^2\propto\omega^2_m$ grows so much, that IPs are almost merge, i.e., the higher-order sidebands barely have enough power to provide the insensitivity.

Considering Fig.~\ref{Exp2}, the absorption of the first sidebands grows from $15$\% to $35$\% for the corresponding temperature growth from $45$ to $65$~$^{\circ}$C. At a higher temperature, the first sidebands are absorbed to a greater extent, and the smaller RF power is required to achieve IPs. %It appears that under the experimental conditions the shift of IPs frequencies from these of PZDs is of the order of $-20$~Hz for $65$~$^{\circ}$C, if we neglect its value for $45$~$^{\circ}$C.
The change of IPs frequencies is mostly given by the temperature-induced shift from the buffer gas~\cite{doi:10.1063/1.331467} (under the experimental conditions the spin-exchange shift can be neglected~\cite{PhysRevA.73.033414}). Thus, the shift of IPs frequencies from these of PZDs is no more than $-1$~Hz for $65$~$^{\circ}$C, if we neglect its value for $45$~$^{\circ}$C.

Dependence of the error-signal frequency on $\omega_m$ can be suppressed by reducing the $\mathcal{K}$ value via adjusting the detuning $\Delta_L$. From the steady-state solution of Eqs.~\eqref{initialequations} follows that the CPT resonance is symmetrical and the responses are odd functions of $\tilde{\delta}$ in the case $\mathcal{K}=0$, even if $\mathcal{E}^2_{-1}\neq\mathcal{E}^2_1$. As a consequence, the zero-crossing point is not shifted by $\delta_{as}$. The equality $\mathcal{K}=0$ can be fulfilled at a certain value of the one-photon detuning~\cite{1418443}, which we call here the symmetrizing value. For $^{87}$Rb atoms it is determined (in MHz) by the following condition:
\begin{equation}
\frac{3\Delta_L}{\Delta^2_L+\Gamma^2}+\frac{\Delta_L+817}{(\Delta_L+817)^2+\Gamma^2}=0,
\label{symmetrizingdetuning}
\end{equation}

\noindent showing that the symmetrizing value depends on parameter $\Gamma$ and lays somewhere between the excited-state levels. Herewith, as follows from Eq.~\eqref{symmetrizingdetuning} and fitting of the unresolved optical line by the Lorentzians ($\Gamma^2/(\Delta^2_L+\Gamma^2)+(1/3)\Gamma^2/[(\Delta_L+\omega_e)^2+\Gamma^2]$), the difference between the pulled and symmetrizing values of $\Delta_L$ decreases with growth of $\Gamma$ value. For example, at $\Gamma/2\pi$ equal to $1$ and $3$~GHz ($\sim50$ and $\sim150$~Torr of the Ar and N$_2$ mixture) the value of parameter $\Gamma\mathcal{K}/(V_{L_u}V_{R_u})$ is $\sim-0.11$ and $\sim-0.02$ at the pulled values of $\Delta_L/2\pi$ equal to $-87$ and $-52$~MHz, respectively.

Therefore we have studied two additional atomic cells with $44$ and $66$~Torr of the buffer gas pressure.
The results are summarized in Table~\ref{Table3}. The obtained frequency values of the CPT resonance for $\omega_m/\tilde{\Gamma}_g=1/2,1$ clearly show that their difference becomes smaller as pressure grows and, as a consequence, $\Delta_L$ gets closer to the symmetrizing value. The frequency shift between curves was reduced by $\sim1.5$ ($\sim5$) times under change from $19$ to $44$ ($60$)~Torr, since the displacement was reduced from $\sim10.5$ to $\sim7$ ($\sim2$)~Hz. Therefore the corresponding tuning of the laser frequency can be used to reduce the effect of sidebands asymmetry on the error-signal frequency shift.
\begin{table}[ht]
\caption{Frequency of the CPT resonance $\omega_R/2\pi$\textbackslash error, Hz. The values are taken for the minimums of the frequency dependencies on the modulation power.}
\begin{tabularx}{\columnwidth}{ 
  | >{\centering\arraybackslash}X 
  | >{\centering\arraybackslash}X 
  | >{\centering\arraybackslash}X | } 
 \hline
 Pressure,~Torr & $\omega_m/\tilde{\Gamma}_g=1/2$ & $\omega_m/\tilde{\Gamma}_g=1$ \\
 \hline
 19 & 3061.5\textbackslash0.3 & 3051\textbackslash0.3 \\
 \hline
 44 & 7090\textbackslash0.5 & 7083\textbackslash0.5 \\
 \hline
 60 & 10851\textbackslash0.5 & 10849\textbackslash0.5 \\
 \hline
\end{tabularx}
\label{Table3}
\end{table}

\section{Discussion}

Ideally, when the total laser intensity, distribution of optical power between spectral components, parameters $a$, $\omega_m$, temperature, do not drift, we have the stable clocks, and divergence between frequencies of IPs and PZDs is irrelevant. While the modulation parameters $a$, $\omega_m$, and temperature of the atomic cell can be controlled with good enough levels, the VCSELs undergo so-called aging and their optical field changes. Unfortunately, there are no studies in the literature demonstrating which parameters change in the first place. If the distribution changes, then the dependence minimum of the error-signal frequency on the RF power can be chosen as the working point. If the total power changes, then either the first IP is used, or the value of $\omega_m$ is finely tuned to move two first IPs to the minimum, see the curve for $\omega_m/2\pi=1200$~Hz in Fig.~\ref{Exp1}. If the change in the ratio of the first sidebands powers is more critical, then the laser frequency can be adjusted~\cite{Zhang:16} to reduce $\delta_{as}$ value, and, consequently, the divergence between frequencies of IPs and PZDs.

We note that the fine-tuning approach requires a fixed ratio of the first sidebands powers. Otherwise, the frequency of the minimum will change with the drift of this ratio, despite its linear dependencies on total power, and the RF power are suppressed. Therefore, at least the clocks frequency reproducibility will be decreased.

\section{Summary}

We have demonstrated that there are two different conditions on the laser spectra, providing the insensitivity to changes in the optical field intensity and the zero displacement of the error-signal frequency from that of the ``$0-0$'' transition. The divergence arises due to the inequality of the first sidebands powers or optically thick medium, making the frequency shift a nonlinear function of the laser power. The first source of the effect makes it more difficult to reach the insensitivity compared to the linear case, i.e., deeper modulation of the VCSEL injection current is required. In contrast, the increase of the first sidebands absorption has the opposite effect, i.e., the insensitivity can be reached at smaller modulation depths. At the same time, both sources result in a decreased degree of the insensitivity. Namely, the linear response of the error-signal frequency can be reduced only to a quadratic one. Whereas in the optically thin medium at equal powers of the first sidebands, the suppression both of the displacement and sensitivity can be total.

We have also shown that the error-signal frequency shift from inequality of the first sidebands powers depends on the laser frequency and decreases when the laser frequency approaches the so-called symmetrizing value. This allows us to conclude that locking of the laser frequency to the vicinity of the symmetrizing value (via stabilization to the slope of the absorption contour) can be useful when the ratio of the resonant sidebands powers varies with the VCSEL aging. On the other hand, if the change of the total power is more critical, then the difference in the first sidebands powers can be used positively. It allows the fine tuning, i.e., the minimum of the error-signal frequency dependence on the RF power can be made less sensitive to variations in the total laser power. Also, the pulled value of the one-photon detuning approaches to the symmetrizing one with growth of the homogeneous broadening of the optical line. Thus, the usage of atomic cells with relatively high buffer gas pressure can be helpful to reduce the resonance's asymmetry. Besides, it increases with the growth of the difference between populations of magneto-insensitive sublevels. Therefore, the reduction of the laser intensity can be also useful to make the resonance more symmetric.

If one relies on its modulation to seek a spectrum providing the zero light shift, then it is important to have the error-signal frequency shift as close as possible to a linear function of the optical field power. In the case of nonlinearity, the error-signal frequency will be displaced from that of the ground-state transition unperturbed by optical fields. Moreover, the displacement can undergo drifts (for example, due to a change in the ratio of the first sidebands powers), which will worsen the long-term frequency stability and reproducibility of an atomic clock. The criterion of the linearity is the coincidence of the error-signal frequency for different spectra providing the insensitivity to variations in the laser power.

\section{Acknowledgments}

The work is supported by the Russian Science Foundation (grant No. 19-12-00417).

\bibliographystyle{apsrev4-1}
\bibliography{references}

\end{document}